\documentclass[aps,pra,twocolumn,floatfix,superscriptaddress]{revtex4}
\usepackage{pstricks,graphicx,amsmath,bbm,mathrsfs,amssymb,siunitx,times}
\newcommand{\ket}[1]{\vert #1 \rangle} \newcommand{\bra}[1]{\langle #1 \vert}

\def\tvarphi{\hat{\varphi}}
\def\var{{\rm var}}
\begin{document}
\title{Squeezing phase diffusion}
\author{Simone Cialdi}
\affiliation{Dipartimento di Fisica ``Aldo Pontremoli'', Universit\`a degli Studi 
di Milano, I-20133 Milano, Italia}
\affiliation{INFN, Sezione di Milano, I-20133 Milano, Italia}
\author{Edoardo Suerra}
\affiliation{Dipartimento di Fisica ``Aldo Pontremoli'', Universit\`a degli Studi 
di Milano, I-20133 Milano, Italia}
\affiliation{INFN, Sezione di Milano, I-20133 Milano, Italia}
\author{Stefano Olivares}
\affiliation{Dipartimento di Fisica ``Aldo Pontremoli'', Universit\`a degli Studi 
di Milano, I-20133 Milano, Italia}
\affiliation{INFN, Sezione di Milano, I-20133 Milano, Italia}
\author{Stefano Capra}
\affiliation{Dipartimento di Fisica ``Aldo Pontremoli'', Universit\`a degli Studi 
di Milano, I-20133 Milano, Italia}
\affiliation{INFN, Sezione di Milano, I-20133 Milano, Italia}
\author{Matteo G.~A.~Paris}
\affiliation{Dipartimento di Fisica ``Aldo Pontremoli'', Universit\`a degli Studi 
di Milano, I-20133 Milano, Italia}
\affiliation{INFN, Sezione di Milano, I-20133 Milano, Italia}
\date{\today}
\begin{abstract}
We address the use of optical parametric oscillator (OPO) to 
counteract phase-noise in quantum optical communication 
channels, and demonstrate reduction of phase diffusion 
for coherent signals travelling through a suitably tuned OPO. 
In particular, we theoretically and experimentally show that 
there is a threshold value on the phase-noise, above which OPO 
can be exploited to ``squeeze'' phase noise. The threshold 
depends on the energy of the input coherent state, and on 
the relevant parameters of the OPO, i.e. gain and input/output 
and crystal loss rates.
\end{abstract}
\maketitle
\section{Introduction}\label{s:intro}
The encoding of information onto the phase of an optical
signal represents a basic building block for quantum enhanced 
metrology and communication  \cite{caves, holland, giovannetti:1, giovannetti:2, demko, escher, chaves, smirne, dary}.
In particular, protocols based on coherent phase-shift-keyed 
signals are useful in those scenarios where single photons
and entanglement may not be the optimal choice, as it happens 
in free-space communication. In those cases, the major obstacle 
to fully exploit the advantages of quantum measurements, thus 
beating the shot-noise limit, is phase noise due to phase 
diffusion  \cite{bina, witt, muller, izumi, genoni, witt2, tsuj}.
\par
Besides trying to avoid phase noise, a question thus arises on
whether, and how, it may be possible to contrast and counteract  
the effects of phase diffusion, and to decrease its detrimental 
effect on the coherence of the signal. In this framework, a 
natural choice to hamper phase diffusion is to use phase-sensitive
amplification, as that provided by optical parametric oscillators 
(OPOs) in the degenerate regime. Despite the simplicity of the 
underlying idea the use of OPO to {\em squeeze} phase diffusion
did not receive much attention in the past. The reason is twofold: 
on the theoretical side, the existence of quantum limits to 
amplification suggests that no full compensation of noise 
is possible \cite{qlm1,qlm2,qlm3,qlm4,qlm5}.
On the experimental side, {\em seeding} a quantum amplifier
with a phase diffused state in order to compensate the noise is
not a trivial task. Likewise, achieving the necessary level of 
pump phase stabilization may be challenging. 
\par
In this paper, thanks to an active stabilisation scheme of 
our OPO cavity and to a novel technique for pump stabilisation, 
we have been able to explore experimentally the use of optical 
parametric oscillator (OPO) to counteract phase-noise, and to 
demonstrate reduction of phase diffusion for coherent signals.
In particular, we theoretically and experimentally show that 
there is a threshold value on the phase-noise, above which 
OPOs may be exploited to effectively {\em squeeze} phase noise. 
As we will see, the noise threshold depends on the amplitude of 
the input signal, and on the relevant parameters of the OPO, 
i.e. gain and input/output and crystal loss rates.
\par
The paper is structured as follows. In Section \ref{s:theory}, 
we introduce notation and discuss the effect of an OPO on a 
phase-diffused coherent signal. In Section \ref{s:experiment}, 
we briefly describe our experimental setup and present our 
results, showing that above a given threshold of the phase 
diffusion amplitude, OPO may be indeed exploited to squeeze 
phase noise. Section \ref{s:conclusions} closes the paper 
with some concluding remarks.
\section{Phase diffusion and OPO}\label{s:theory}
\begin{figure}[h!]
\includegraphics[width=0.95\columnwidth]{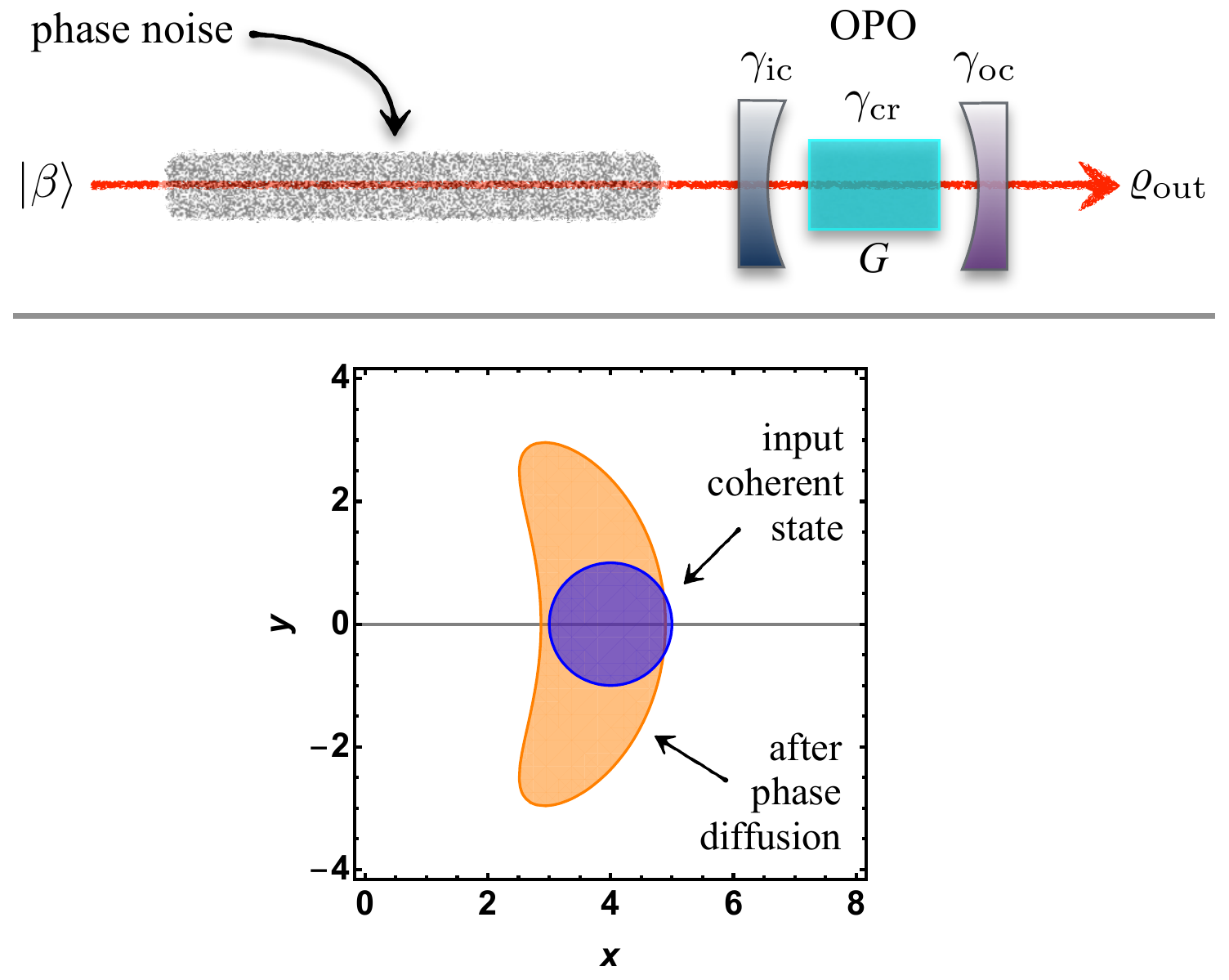}
\vspace{-0.3cm}
\caption{(Upper panel):
After the phase diffusion the noisy signal enters an OPO 
characterised by the gain $G$ and by the input, $\gamma_{\rm ic}$,
output, $\gamma_{\rm oc}$ and crystal $\gamma_{\rm cr}$ loss rates. 
(Lower panel): phase-space representation a coherent state 
$\ket{\beta}$ (blue) and
its phase diffused counterpart (orange). We set $\beta = 2$ and 
$\sigma = \pi/4$.
It is clear the effect of phase noise on the uncertainty of the $x$ and $y$ quadratures.}\label{f:fig:PhDiff}
\end{figure}

Let us consider a coherent state $\ket{\beta \, e^{i\varphi}}$, 
with $\beta \in {\mathbbm R}_{+}$ undergoing phase diffusion. 
The evolved state may be written as \cite{genoni,geno:12}
\begin{equation}
\label{ph:diff:ch}
\varrho = \int\!\! d\phi\, g_\sigma(\phi)\;\ket{\beta\,e^{i\phi}}\bra{\beta
\,e^{i\phi}}\,,
\end{equation}
where $g_\sigma(\phi) = (2\pi\sigma^2)^{-\frac12}\exp\{-\frac12 \phi^2/\sigma^2\}$ and we refer to $\sigma$ as to the amplitude 
of the phase diffusion (see Fig. \ref{f:fig:PhDiff}).
An estimate of the phase $\varphi$ may be obtained
from the expectations of the two orthogonal quadratures 
$x = a + a^\dag$ and $y=i(a^\dag - a)$, where
$a$ and $a^\dag$ are the creation and annihilation 
operators, $[a,a^\dag]=1$
\begin{equation}\label{est:phi}
\tvarphi = \tan^{-1} \frac{\langle y \rangle}{\langle x 
\rangle}\,.
\end{equation}
The uncertainty in the estimate $\tvarphi$ is thus given by
namely:
\begin{align}\label{var:phi}
\var[\tvarphi] = & \left(\frac{\langle y \rangle}{\langle x \rangle^2 + \langle y \rangle^2}\right)^2 \var[x] \notag \\ 
& +\left(\frac{\langle x \rangle}{\langle x \rangle^2 + \langle y \rangle^2}\right)^2 \var[y]\,.
\end{align}
For a coherent signal $\var[x] = \var[y] = 1$ 
\begin{equation}
\var[\tvarphi] = \frac{1}{4\, \beta^2}\,,
\end{equation}
that scales, as expected, as the inverse of the energy 
$|\beta|^2$ of the input coherent state. Notice
that we estimate phase by two measurements
performed on two copies of the input state \cite{oli:rev} 
(thanks to stability of the scheme). 
Let us now consider a coherent signal undergoing phase
diffusion, and for the sake of simplicity let us assume 
that $\varphi = 0$, i.e., $\beta \in {\mathbbm R}$ and 
$\beta >0$. In this case we have:
\begin{align}
\langle x \rangle = 2 |\beta| e^{-\sigma^2} \quad \mbox{and}\quad
\langle y \rangle = 0 \, ,
\end{align}
and:
\begin{align}
\var[x] &= 1 + 2 \beta^2 \left(1 - e^{-\sigma^2}\right)^2\,, \\[1ex]
\var[y] &= 1 + 2 \beta^2 \left(1 - e^{-2\sigma^2}\right)\,.
\end{align}
Note that $\var[x] \le \var[y]$. It is straightforward to show that:
\begin{equation}\label{var:Phdiff}
\var[\tvarphi] = \frac{\cosh(\sigma^2) + (1+4 \beta^2) \sinh(\sigma^2)}{4\beta^2} \ge \frac{1}{4\,\beta^2}
\end{equation}
In Fig.~\ref{f:fig:PhDiff} we compare a coherent state with its phase diffused counterpart.
\par
Let us now assume that after the phase diffusion process the 
degraded state passes through an OPO as sketched in the 
upper panel of Fig.~\ref{f:fig:PhDiff}. The OPO can be
 characterised by the gain and the input and output parameters,
$\eta_{\rm in}$ and $\eta_{\rm esc}$, respectively, which summarise the effect of the input and output couplers
transmissivity and the internal losses \cite{bachor}. The gain can be written as $G = (1-d)^{-2}$, where $d = \sqrt{P/P_{\rm th}}$,
where $P$ is the pump power and $P_{\rm th}$ the OPO power threshold; the input and output parameters depends
on the input ($\gamma_{\rm ic}$) and output ($\gamma_{\rm oc}$) loss rates of the input and output couplers, respectively,
but also on the loss rate due to the OPO crystal ($\gamma_{\rm cr}$). Overall, we have:
\begin{equation}
\eta_{\rm in} = \frac{\gamma_{\rm ic}}{\gamma}
\quad\mbox{and}\quad
\eta_{\rm esc} = \frac{\gamma_{\rm oc}}{\gamma}\,,
\end{equation}
where we introduced $\gamma = \gamma_{\rm ic}+\gamma_{\rm oc}+2\gamma_{\rm cr}$.
In the following we will focus on the effect of the OPO on the first and second moments of the quadratures
relevant for our analysis.
\begin{figure}[htb!]
\centering
\includegraphics[width=0.98\columnwidth]{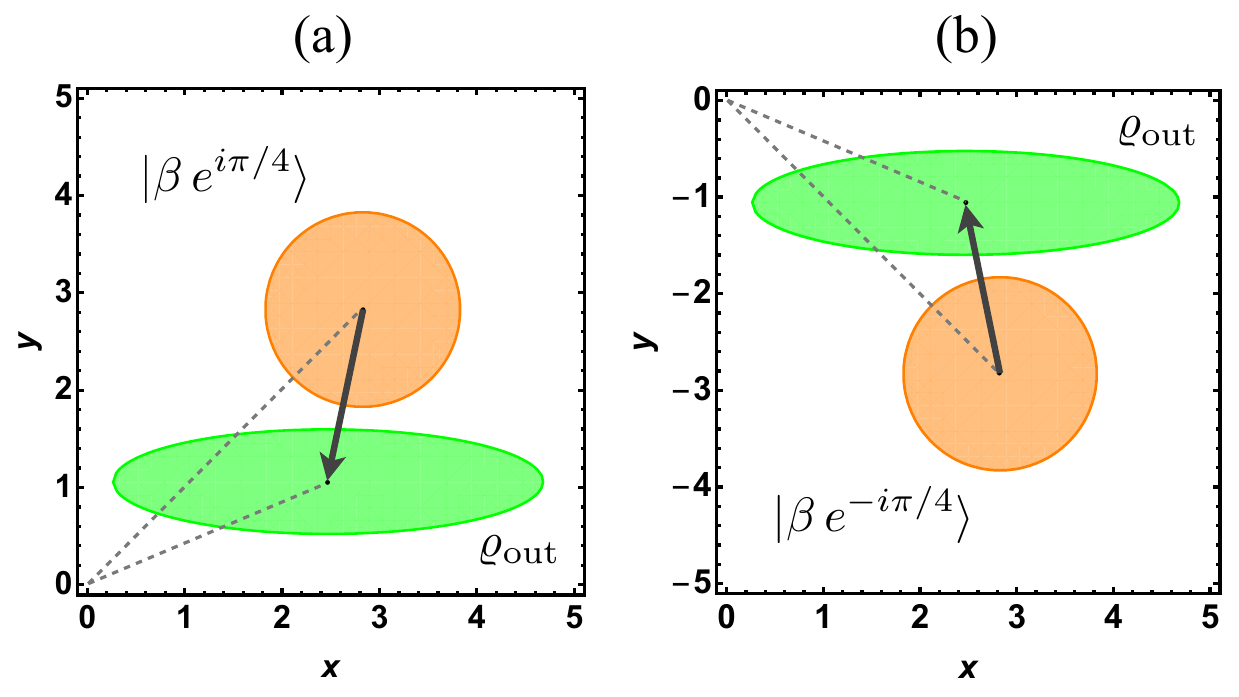}
\vspace{-0.3cm}
\caption{Phase-space representation a coherent $\ket{\beta\, e^{i\varphi}}$ state before (blue) and
after the evolution through the OPO (green). We set $\beta = 2$ and two values of the
phase: (a) $\varphi = \pi/4$ and (b) $\varphi = -\pi/4$.
In both the cases the phase approaches $0$ after the evolution.
We used the realistic OPO parameters $d=0.40$ ($G = 2.78$), $\eta_{\rm in} = 0.08$ and $\eta_{\rm esc} = 0.87$.
}\label{f:fig:OPO}
\end{figure}
We begin our study assuming as input a coherent state $\ket{\beta \, 
e^{i\phi}}$ (without phase noise).
Lengthy but straightforward calculations lead to the following results
(we choose the OPO pump phase such to amplify the $x$ quadrature):
\begin{subequations}
\label{ave:OPO}
\begin{align}
X &\equiv \langle x \rangle = \frac{\sqrt{4\,\eta_{\rm in}\eta_{\rm esc}}}{1-d}\, 2  \beta \cos \varphi\,,\\[1ex]
Y &\equiv \langle y \rangle = \frac{\sqrt{4\,\eta_{\rm in}\eta_{\rm esc}}}{1+d}\, 2 \beta \sin \varphi\, ,
\end{align}
\end{subequations}
and
\begin{subequations}
\label{var:OPO}
\begin{align}
\Sigma^2_x (\eta_{\rm esc},d) &\equiv \var[x] = 1 + \eta_{\rm esc} \frac{4d}{(1-d)^2} \ge 1\,, \\[1ex]
\Sigma^2_y (\eta_{\rm esc},d) &\equiv \var[y] = 1 - \eta_{\rm esc} \frac{4d}{(1+d)^2} \le 1\,.
\end{align}
\end{subequations}
In Fig.~\ref{f:fig:OPO} we show the effect of the OPO on a coherent state for a particular choice
of the involved parameters. It is worth noting that the presence of the OPO reduces the phase shift
of the state: this effect will be fundamental to reduce the phase diffusion.

\begin{figure}[htb!]
\centering
\includegraphics[width=0.98\columnwidth]{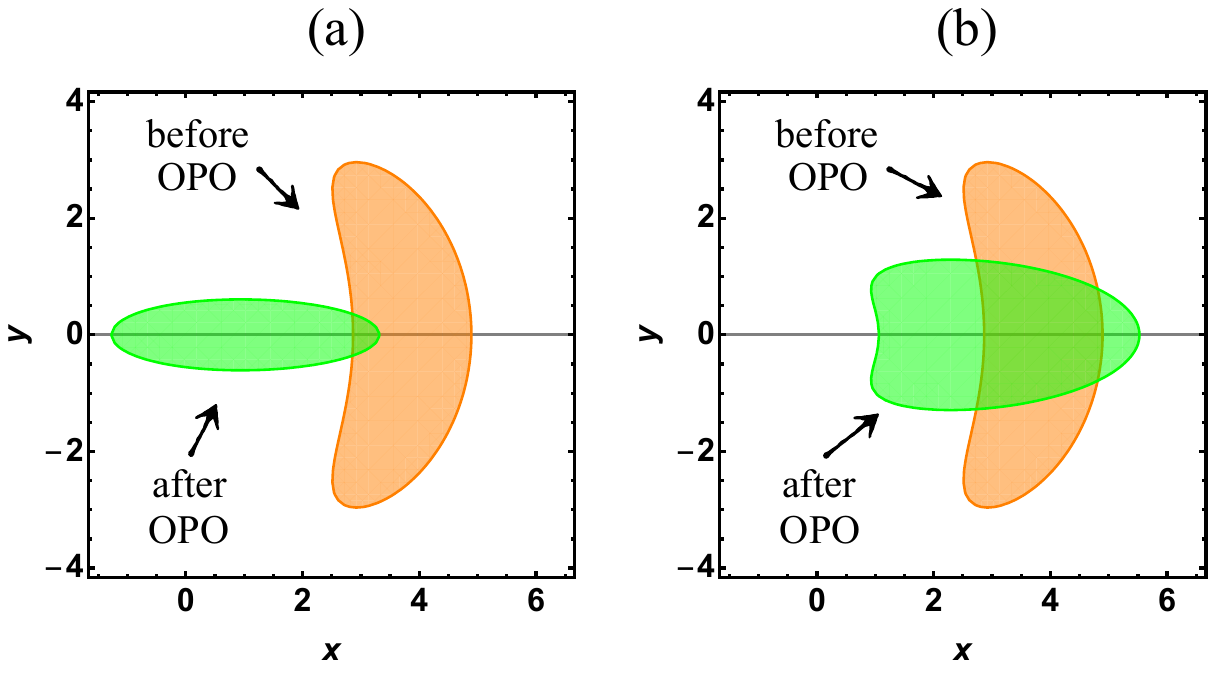}
\vspace{-0.3cm}
\caption{Phase-space representation of a phase diffused coherent state (orange) and
its state $\varrho_{\rm out}$ after the evolution through the OPO (green).
We set $\beta = 2$, $\varphi=0$ for the input coherent state and $\sigma=\pi/4$ for the
phase noise amplitude and $d=0.4$ ($G = 2.78$). For the input and output parameters we used
the realistic values (a)~$\eta_{\rm in} = 0.01$, $\eta_{\rm esc} = 0.93$ and (b) $\eta_{\rm in} = 0.08$, $\eta_{\rm esc} = 0.87$.
We can see that the uncertainty of the $y$ quadrature is reduced after the evolution through the OPO.}\label{f:fig:PhDiff:OPO}
\end{figure}
When phase noise affects the propagation of the coherent state before the OPO, the evolved state
is given by Eq.~(\ref{ph:diff:ch}). By using the results of Eqs.~({\ref{ave:OPO}}) and (\ref{var:OPO})
it is easy to find the new mean values of the relevant quadratures and their variances, namely
(we set $\varphi = 0$):
\begin{subequations}
\label{ave:OPO:PhDiff}
\begin{align}
\langle x \rangle = \alpha_x \, e^{-\sigma^2/2}\, \qquad
\langle y \rangle = 0\, ,
\end{align}
\end{subequations}
and
\begin{subequations}
\label{var:OPO:PhDiff}
\begin{align}
\var[x] &= \Sigma^2_x + \alpha_x^2\, e^{-\sigma^2} \left[ \cosh (\sigma^2) - 1 \right]\,, \\[1ex]
\var[y] &= \Sigma^2_y + \alpha_y^2\, e^{-\sigma^2} \sinh (\sigma^2) \,,
\end{align}
\end{subequations}
where:
\begin{subequations}
\begin{align}
\alpha_x &\equiv \alpha_x(\beta, \eta_{\rm in}, \eta_{\rm esc}, d) = \frac{\sqrt{4\,\eta_{\rm in}\eta_{\rm esc}}}{1-d}\, 2  \beta\,,\\[1ex]
\alpha_y &\equiv \alpha_x(\beta, \eta_{\rm in}, \eta_{\rm esc}, d) =\frac{\sqrt{4\,\eta_{\rm in}\eta_{\rm esc}}}{1+d}\, 2 \beta\,,
\end{align}
\end{subequations}
whereas $\Sigma^2_x$ and $\Sigma^2_y$ are given in Eqs.~({\ref{var:OPO}}).
In Fig.~\ref{f:fig:PhDiff:OPO} we can see how a phase diffused coherent state is modified
by the evolution through the OPO.
\begin{figure}[htb!]
\centering
\includegraphics[width=0.8\columnwidth]{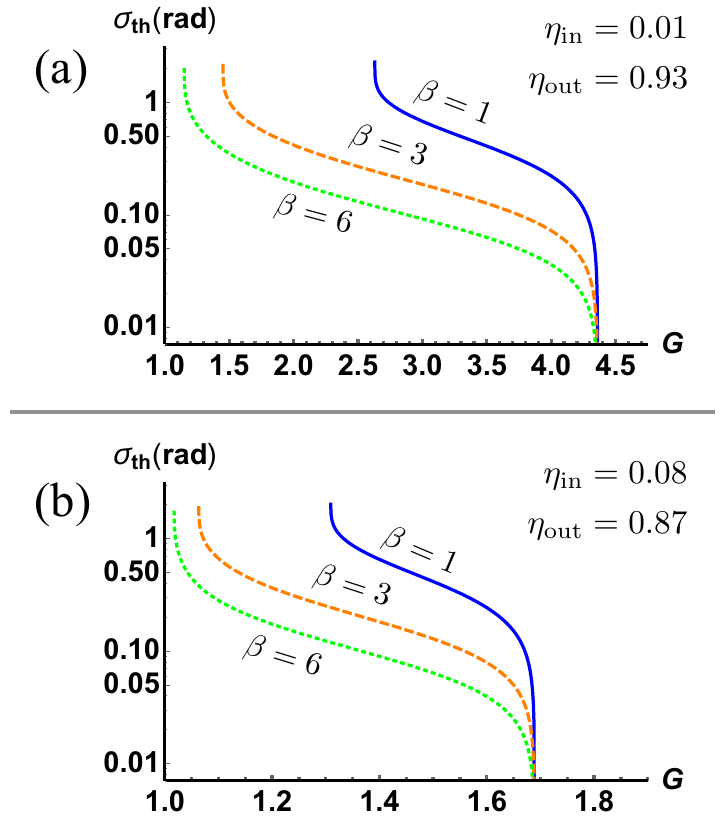}
\vspace{-0.3cm}
\caption{Threshold value $\sigma_{\rm th}$ of the phase noise as a function of the OPO gain $G = (1-d)^{-2}$
for different values of the input coherent state $\beta$: if $\sigma > \sigma_{\rm th}$ the OPO reduces the phase noise.
The input and output parameters of the panels are $\eta_{\rm in} = 0.01$ and $\eta_{\rm esc} = 0.93$ (top panel) and
$\eta_{\rm in} = 0.08$ and $\eta_{\rm esc} = 0.87$ (right panel).
}\label{f:fig:sigmath}
\end{figure}
In order to assess the reduction of phase diffusion, we use the previous results to
evaluate $\tvarphi$ and $\var[\tvarphi]$ given in Eqs.~(\ref{est:phi}) and (\ref{var:phi}).
In particular, for the variance we find:
\begin{equation}\label{var:OPO:Phdiff}
 \var[\tvarphi] = \frac{\Sigma^2_y + \alpha_y^2\, e^{-\sigma^2} \sinh(\sigma^2)}{\alpha_x^2\, e^{-\sigma^2} }\,.
\end{equation}
By comparing the variance (\ref{var:OPO:Phdiff}) and the one obtained without the OPO, namely, Eq.~(\ref{var:Phdiff}), we can
find a threshold value $\sigma_{\rm th}$ of the phase noise amplitude above which phase diffusion can be reduced, namely:
\begin{align}\label{eq:threshold}
&\sigma_{\rm th}(\beta, \eta_{\rm in}, \eta_{\rm esc}, d) =\nonumber\\
&\hspace{1cm}
\sqrt{
\log\left[
\frac{\sqrt{2}\beta\sqrt{\alpha_x^2 - \alpha_y^2}}{\sqrt{\alpha_x^2 + 2 \beta^2 (\alpha_x^2 - \alpha_y^2 - 2\Sigma_x^2)}}
\right]
}\,.
\end{align}
In Fig.~\ref{f:fig:sigmath} we plot the threshold $\sigma_{\rm th}$ as a function of the OPO gain $G = (1-d)^{-2}$
for different values of the involved parameters. We can see that, for a fixed choice of the involved parameters,
there exists a maximum value of $G$ above which the OPO always reduces phase noise.
\section{Experimental results}\label{s:experiment}
\begin{figure}[h!]
\centering
\includegraphics[width=0.9\columnwidth]{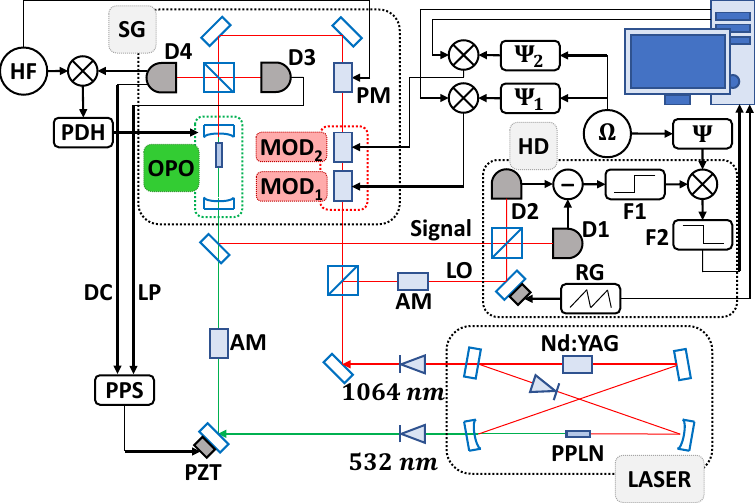}
\vspace{-0.3cm}
\caption{Schematic diagram of the experimental setup. The main source is the Nd:YAG laser internally frequency-doubled, while OPO is in the state generation stage (SG). States are revealed by a homodyne detector (HD). Generation and detection are fully controlled by a computer.
}\label{f:setup}
\end{figure}

Theoretical predictions have been tested using the experimental scheme depicted in Fig.~\ref{f:setup}. The setup allows us to generate and manipulate displaced-squeezed states.
In particular, we have control not only on both the amplitude and phase of the states, but also on the gain $G$ and the pump phase.
The setup is described in detail in \cite{mandarino:16} and consists of different stages: a Laser source (LASER), a state generation and manipulation stage (SG) and an homodyne detector (HD).
A home-made \SI{1064}{\nano\meter} wavelength Nd:YAG laser internally frequency doubled at \SI{532}{\nano\meter} serves both as the input seed and the pump beam for the Optical Parametric Oscillator (OPO). In particular, from the infrared output of the laser we generate two different beams by a polarising beam splitter: one is used as the local oscillator (LO) for the homodyne detection, while the other is sent to the SG.
\par
OPO consists in a linear cavity with a free spectral range of $\SI{3.270}{\giga\hertz}$.
A $\SI{10}{\milli\meter}$-long MgO:LiNbO3 crystal with anti-reflection coating is inserted inside the cavity. The losses related to the crystal are $\Delta=2.42\times 10^{-3}$.
We used two different configurations A and B for the OPO.
In configuration A the input mirror has a reflectivity $R_{\rm ic,A}=0.999$ with a radius of curvature of $\SI{10}{\milli\meter}$, while the output mirror has a reflectivity $R_{\rm oc}=0.917$ with a radius of curvature of $\SI{25}{\milli\meter}$, leading to $\eta_{\rm in,A}=0.008$ and $\eta_{\rm esc,A}=0.937$, with a measured cavity total transmissivity $T_{\rm A}=0.029$.
In configuration B the input mirror has a reflectivity $R_{\rm ic,B}=0.9925$, and the output mirror is the same as A, both with the same radii of curvature of A, leading to $\eta_{\rm in,B}=0.079$ and $\eta_{\rm esc,B}=0.871$, with a measured cavity total transmissivity $T_{\rm B}=0.26$.
The role of the transmissivities will be clear later.
The cavity is actively stabilized using the Pound-Drever-Hall (PDH) technique \cite{cialdi:16} by means of a phase modulator (PM) placed along the SG beam, which generates two \SI{116}{\mega\hertz} sidebands around the laser frequency.
Coherent states are generated exploiting the combined effect of two optical modulators (MOD1 and MOD2) placed before the OPO. A proper choice of their modulations allows us to generate an arbitrary coherent state on the sidebands at \SI{3}{\mega\hertz} for seeding the OPO \cite{mandarino:16,cialdi:16}. Amplitude and phase values are set on demand by a computer.
\par
In order to effectively amplify a certain quadrature with the OPO, the pump phase $\Theta$ must be stable over the whole time of the measurement. Therefore, a novel technique for the pump phase stabilization (PPS) has been developed and will be
described in detail in a forthcoming work \cite{bib:cialdi2030}, whereas in the following we highlight its main elements.
In our stabilization technique, the field $E_{\rm r\,tot}$ reflected by the OPO can be exploited as an error signal for the stabilization of $\Theta$. We note that the field $E_{\rm r\,tot}$ is the sum of the field $E_{\rm r}$, directly reflected by the input mirror, and the field $E_{\rm t}$ transmitted by the cavity through the input mirror.
The laser and the field generated by down conversion by the pump interact inside the cavity and their interaction leads to constructive or destructive interference depending on the pump phase $\Theta$.
Since $E_{\rm r}$ and $E_{\rm t}$ are $\pi$-shifted, in the first case $E_{\rm t}$ increases and $E_{\rm r\,tot}$ decreases, while in the second case $E_{\rm t}$ decreases and $E_{\rm r\,tot}$  increases.
This can be summarized as $E_{\rm r\,tot}=E_{\rm r\,tot}(\Theta)$, thus in order to access the information about $\Theta$ we measure the corresponding power $P_{\rm r tot}$ with the DC output of the detector $\rm D_{\rm 4}$ (see Fig.~\ref{f:setup}).
%
Finally, the obtained error signal is properly manipulated with a PID and applied to a piezoelectric actuator which can change $\Theta$.
At the moment the total bandwidth of the PPS is around \SI{1}{\kilo\hertz}, thus we can compensate pump phase fluctuations of at most that frequency.
\par
\begin{figure}[t!]
\includegraphics[width=0.495\columnwidth]{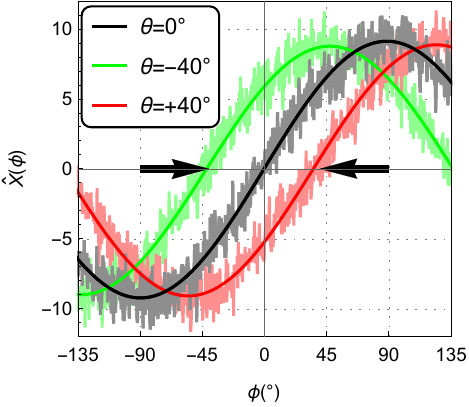}
\includegraphics[width=0.495\columnwidth]{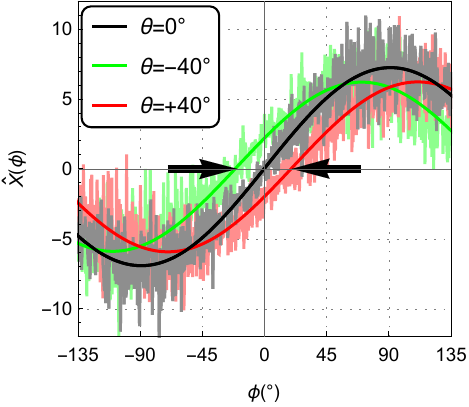}
\includegraphics[width=0.495\columnwidth]{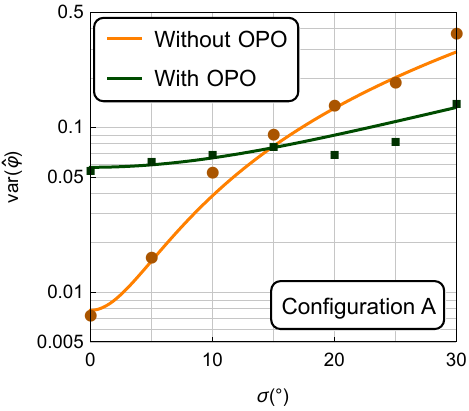}
\includegraphics[width=0.495\columnwidth]{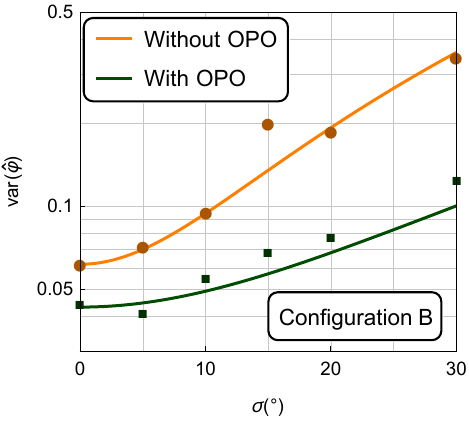}
\vspace{-0.7cm}
\caption{
(Upper panels):
Tomography of a sequence of three \SI{40}{\degree}-shifted coherent states (left) and relative effect of OPO (right). The marked line is the mean value of each state. Notice the reduction of the phase shift as marked by the arrows. Here the system is in configuration A, but the amplitude of the states is different in the two cases without and with OPO ($\beta=4.5$ and $\beta_{\rm OPO}=2.0$ with $G=3.1$, respectively).
(Lower panels):
Variance of $\varphi$ in function of the phase diffusion for the configurations A (left) and B (right). Theoretical curves (lines) fit well the experimental points. Notice the presence of the threshold at \SI{14.8}{\degree} in configuration A.
\label{f6}}
\end{figure}
The effect of the OPO on a phase-diffused coherent state has been tested directly by generating a rapid sequence of three coherent states: one with $0$-phase shift and two with a $\pm\SI{40}{\degree}$-phase shift.
Indeed, considering the Eqs.~({\ref{ave:OPO}}), if the phase shift before OPO is $\varphi = \theta_0$, then after OPO we have
\begin{equation}
\frac{\langle y \rangle}{\langle x \rangle}=\frac{1-d}{1+d}\tan{\varphi},
\end{equation}
leading to the simple relation
\begin{equation}
\label{p-squeez}
\tan{\theta_d}=\frac{1-d}{1+d}\,\tan{\theta_0}
\end{equation}
where $\theta_d$ and $\theta_0$ refer to the case with and without OPO respectively.
We performed a tomography of every state, both with and without the OPO. Results are shown in the upper panels of Fig.~\ref{f6}.
In order to highlight the effect of the OPO on the phase shift, we used two different amplitudes for the cases with and without the OPO and this is possible as the reduction of the phase shifts does not depend on the $\beta$s [it is clear from Eq.~(\ref{p-squeez})].
In particular the system was in configuration A, with $\beta=4.5$ without OPO and $\beta_{\rm OPO}=2.0$ with OPO. In this last case the gain was $G=3.1$, leading to $d=0.43$.
Results show that the phase shift is reduced from $\theta_0=\SI{40}{\degree}$ to a measured $\theta_{d,\rm ex}=\SI{20}{\degree}$, while the theoretical value calculated with $d$ is $\theta_{d,\rm th}=\SI{18.4}{\degree}$.
\par
In our experiments, we generated the phase-diffused coherent states by modulating their phases with a suitable Gaussian distribution \cite{oli:2013} and we evaluated $\var[\tvarphi]$ of Eq.~(\ref{var:OPO:Phdiff}) with and without OPO for different values of the phase diffusion $\sigma$ in both configurations A and B.
For practical reasons, in the measurements without OPO we did not physically remove it from the setup, but we increased the amplitude of the coherent state by a factor $1/\sqrt{T}$ with respect to the case of squeezed coherent, in order to compensate the effect of the cavity transmissivity $T$. That is why we have measured transmissivities. Of course, pump was turned off during these measurements.
In order to obtain $\var[\tvarphi]$, we measured $\langle x \rangle$, $\langle y \rangle$, $\var[x]$ and $\var[y]$ and we used  Eq.~(\ref{var:phi}). We made measurements for different values of the phase diffusion $\sigma$.
The theoretical expectation has been obtained directly using Eq.~(\ref{var:OPO:Phdiff}), where we considered the experimental values for $\Sigma_y$, $\beta$ and $G$, from which we calculated $\alpha_x$ and $\alpha_y$. Finally, the threshold $\sigma_{\rm th}$ has been calculated by Eq.~(\ref{eq:threshold}).
Both the experimental points and the theoretical previsions are shown in the lower panels of Fig.~\ref{f6} for the two configurations.
In particular, in configuration A we have $\beta_{\rm A}=5.70$ and $G_{\rm A}=2.75$, with a theoretical threshold $\sigma_{\rm th}=\SI{14.8}{\degree}$ perfectly compatible to experimental points.
In configuration B we have $\beta_{\rm B}=2.05$, $G_{\rm B}=3.12$ and there is no threshold, so the use of the OPO is convenient for every phase diffusion amplidue.



\section{Conclusions}\label{s:conclusions}
In this paper, we have exploited our innovative OPO scheme to address
the use of phase-sensitive amplification to counteract phase-noise 
in quantum phase channels. Our results demonstrate the reduction 
of phase diffusion  for coherent signals travelling through a 
suitably tuned OPO. More generally, we have shown that 
there is a threshold value on the phase-noise, above which OPO 
can be exploited to squeeze phase noise. The threshold 
depends on the energy of the input coherent state, and on 
the relevant parameters of the OPO. Our results may be exploited 
to enhance quantum phase communication channels and for
state preparation in quantum metrological schemes. 
\acknowledgements
This work has been partially supported by FNP TEAM project 
"Quantum Optical Communication Systems" and by UniMI project
PSR2017-DIP-008.

\end{document}